\documentclass[prd,aps,nofootinbib]{revtex4}
\usepackage{amsmath}
\usepackage{graphicx}
\usepackage{amsmath}
\usepackage{amssymb}
\usepackage{latexsym}
\usepackage{color}

\newcommand{\PhiB}{\Phi_{\textsc{DS}}}
\newcommand{\GB}{\mathcal{G}}
\newcommand{\de}{\mathrm{d}}

\begin{document}

\title{Vacuum structure for scalar cosmological perturbations in Modified Gravity Models}

\author{Antonio De Felice} \email{antonio.defelice@uclouvain.be}
\affiliation{Theoretical and Mathematical Physics Group, Centre for
  Particle Physics and Phenomenology, Louvain University, 2 Chemin du
  Cyclotron, 1348 Louvain-la-Neuve (Belgium)}

\author{Teruaki Suyama}
\email{teruaki.suyama@uclouvain.be}
\affiliation{Theoretical and Mathematical Physics Group, Centre for
  Particle Physics and Phenomenology, Louvain University, 2 Chemin du
  Cyclotron, 1348 Louvain-la-Neuve (Belgium)}

\date{\today}

\begin{abstract}
  We have found for the general class of Modified Gravity Models
  $f(R,\GB)$ a new instability which can arise in vacuum for the
  scalar modes of the cosmological perturbations if the background is
  not de Sitter. In particular, the short-wavelength modes, if stable,
  in general have a group velocity which depends linearly in $k$, the
  wave number. Therefore these modes will be in general
  superluminal. We have also discussed the condition for which in
  general these scalar modes will be ghost-like. There is a subclass
  of these models, defined out of properties of the function
  $f(R,\GB)$ and to which the $f(R)$ and $f(\GB)$ models belong, which
  however do not have this feature.
\end{abstract}

\maketitle

\section{Introduction}

Modified Gravity Models (MGM) have been introduced as an alternative
to the quintessence picture for dark energy
\cite{Starobinsky:1980te,Capozziello:2002rd,Carroll:2003wy,Sami:2005zc,Nojiri:2005vv,Carroll:2004de,Dvali:2003rk,DeFelice:2008af,ArkaniHamed:2003uy}. The idea is
speculative but attractive: maybe the acceleration of the universe is
due to a modification of the behaviour of gravity at large
scales. Gravitation seems to be described by an action which is not
renormalizable, and the symmetry of the theory, does not restrict
enough the possibility for building up an action for the $g_{\mu\nu}$
variables, the metric tensor components, as already pointed out in
1917 \cite{kret:1917,einst:1917}. However, to build a sensible
gravitational theory is not an easy task. Some tried to change it to
have a renormalizable theory at 1-loop
\cite{DeWitt:1965jb,Fradkin1982469,Fradkin1981377}, but in general
this leads to fourth-order gravity theories and spurious degrees of
freedom
\cite{Hindawi:1995cu,Hindawi:1995qa,Stelle:1977ry,Barth:1983hb,Boulanger:2000rq,Nunez:2004ts,Chiba:2005nz,Navarro:2005gh}.

Lovelock \cite{Lovelock:1971yv} found and studied a special class of
scalars, the so called Lovelock scalars, built out of the Riemann and
metric tensors only. In four dimensions the only such scalars which
are not identically zero are three: a constant, the Ricci scalar, and
the Gauss-Bonnet combination $\GB$. This last scalar is made of
quadratic combinations of the Riemann tensor, which in turn depends on
second derivatives of $g_{\mu\nu}$. Naively, one might expect that
this term in a Lagrangian would automatically lead to four derivatives
for $g_{\mu\nu}$. However, all the Lovelock scalars have the property
(which makes them special among all possible scalar invariants) that
the equations of motion still remain of second order. This is still
true even if these same curvature invariants are coupled to some other matter scalar
field. Another property of the Gauss-Bonnet term is that, in four
dimensions, it can be written as a total derivative
\cite{Cherubini:2003nj}, therefore the only way for it to give a
contribution to the equations of motion is to couple it to some other
field. Indeed, string theory seems to predict the existence of such
couplings in the low-energy effective action
\cite{Antoniadis:1993jc,Zwiebach:1985uq,Gross:1986mw,Metsaev:1987zx,Easther:1996yd}.

In the aim of building up the action for gravity, since general
covariance is not strong enough, we can take a bottom-up approach and
look for the most general action which can be built out of the Riemann
and the metric tensor. Therefore recently models have been introduced
which use one or both these scalars. These are the so called $f(R)$
\cite{Capozziello:2002rd,Carroll:2003wy,Nojiri:2003ft,Nojiri:2006gh,Capozziello:2006dj,Cognola:2007zu,Capozziello:2008ch,Bazeia:2007jj,Afonso:2007gc,Bazeia:2008ay,Woodard:2006nt,Amendola:2006kh,Amendola:2006we,Song:2006ej,Sawicki:2007tf,Carroll:2006jn,Bean:2006up,Pogosian:2007sw,Laszlo:2007td,Olmo:2005jd,Sotiriou:2008rp},
$f(\GB)$
\cite{Sami:2005zc,Chingangbam:2007yt,Nojiri:2005vv,Nojiri:2005jg,Nojiri:2006je,Koivisto:2006xf,Koivisto:2006ai,Cognola:2006eg,Davis:2007ta,Li:2007jm,Kawai:1998ab,Calcagni:2006ye,DeFelice:2008wz,Zhou:2009cy},
and $f(R,\GB)$ theories
\cite{Carroll:2004de,Nunez:2004ts,Nojiri:2007te,Navarro:2005gh,Navarro:2005da,Mena:2005ta,DeFelice:2006pg}. All
the Lagrangians written through these scalars share the property that
they can be rewritten as scalar fields coupled to $R$ and $\GB$
respectively \cite{DeFelice:2006pg}. Therefore introducing functions
of $R$ and $\GB$ adds scalar degrees of freedom only to the theory,
eliminating the spurious spin-2 extra degrees of freedom which are
typically ghost-like
\cite{Nunez:2004ts,DeFelice:2006pg,Hwang:1999gf,Cartier:2001is}.

It is well known that $f(R)$ theories can be written as a scalar
tensor theory which, through a conformal transformation, takes in the
Einstein frame the same form of the Lagrangian for a scalar field
minimally coupled to gravity
\cite{Teyssandier:1983zz,Wands:1993uu}. However, the same technique
does not prove it useful for the $f(\GB)$ theories. The vacuum structure
of these theories has been deeply studied for maximally symmetric
spaces \cite{Nunez:2004ts,Navarro:2005da}. As for the behaviour of the
same theories on a Friedmann-Lema\^\i tre-Robertson-Walker (FLRW)
background some papers appeared
\cite{Hwang:1999gf,Cartier:2001is,Hwang:2005hb}, but, as we shall see
later, they only treated a subset of the whole class of $f(R,\GB)$
theories.

In this paper we will give a general study of cosmological
perturbations in vacuum for a general $f(R,\GB)$ Lagrangian. Some
papers claimed that they studied the most general MGM
\cite{Cartier:2001is,Hwang:2005hb}. However, we think that only a
subset of these theories was considered (among these, the $f(R)$ and
the $f(G)$ theories). We present here a new result for the most
general MGM, that is the dispersion relation for the scalar
perturbations changes, and typically scalar modes, if stable, will be
superluminal with a speed proportional to the wave number $k$.  This
change for the dispersion relation is always present except for a
special subclass, which actually contains all the models for which
cosmological perturbation theory was discussed
\cite{Cartier:2001is,Hwang:2005hb}. For such models, we recover the
same results found before.

This new feature is essential to study the vacuum structure for these
theories and an additional bound (the speed of the short wavelength
modes) should be added to the previous one found in the literature of
these models. We want to stress out that this is a physical property
of the general class of these theories, and not a feature of a bad
gauge choice, or a spurious feature which can be removed by a suitable
field redefinition. Although the study of this action in the presence
of matter fields is very important, the analysis of the vacuum case is
the first thing to address. We will come back to this point in the
discussion section.

The paper is structured as follows. In section II we introduce the
model together with the equations of motion for a FLRW background. In
section III, we perform a perturbation analysis for the scalar
modes and derive a master equation.  In section IV, we analyze the
master equation and discuss properties of the solution.  In section V,
we consider a toy model of MGM to demonstrate how our findings
can actually set bounds on MGM.  Section VI is devoted to the
study of special cases where the structure of the master equation for
high $k$ modes differs from the general case.  In section VII, based
on our findings obtained from the study of the scalar perturbations,
we discuss various implications which must be taken into account for
building up a sensible MGM.  We also comment on the past studies
where the scalar perturbations for the MGM were studied. Finally, in 
section VIII we present our conclusions.

\section{Modified Gravity Models}

We consider the modified gravity action given by
\begin{eqnarray}
S=\frac{M_P^2}{16\pi} \int d^4x\sqrt{-g}~f(R,\GB), \label{action0}
\end{eqnarray}
where $R$ is the Ricci scalar and $\GB$ is the so-called Gauss-Bonnet
term defined by
\begin{eqnarray}
\GB\equiv R^2-4R_{\mu \nu} R^{\mu \nu}+R_{\mu \nu \alpha \beta} R^{\mu \nu \alpha \beta}.
\end{eqnarray}
Since we want to study the general theory of cosmological
perturbations for the modified gravity action (\ref{action0}), we do
not assume any particular functional form of $f(R,\GB)$.

Although (\ref{action0}) is the basic action we consider, for
convenience of the actual analysis, we mainly use a different action,
equivalent to (\ref{action0}), which is given by
\begin{eqnarray}
S&=&\frac{M_P^2}{16\pi} \int d^4x\sqrt{-g}~\bigg[ f(\lambda,\sigma)+(R-\lambda) F(\lambda,\sigma)+(\GB-\sigma) \xi(\lambda,\sigma) \bigg]\nonumber \\
&=&\frac{M_P^2}{16\pi}\int d^4x\sqrt{-g}~\bigg[ RF(\lambda,\sigma)+\GB\xi(\lambda,\sigma)-V(\lambda,\sigma) \bigg], \label{action1}
\end{eqnarray}
where $\lambda$ and $\sigma$ are auxiliary fields and
\begin{eqnarray}
&&F(\lambda,\sigma) \equiv \frac{\partial f}{\partial \lambda},~~~\xi(\lambda,\sigma) \equiv \frac{\partial f}{\partial \sigma}, \\
&&V(\lambda,\sigma)\equiv \lambda F(\lambda,\sigma)+\sigma\xi(\lambda,\sigma)-
f(\lambda,\sigma).
\end{eqnarray}
By the following way, we can verify that the action (\ref{action1}) is
equivalent to (\ref{action0}).  From the variation of $S$ with respect
to $\lambda$ and $\sigma$, we have equations for $\lambda$ and $\sigma$
which are given by
\begin{eqnarray}
&&(R-\lambda) F_\lambda+(\GB-\sigma) F_\sigma=0, \label{lambda1} \\
&&(R-\lambda) F_\sigma+(\GB-\sigma) \xi_\sigma=0, \label{sigma1}
\end{eqnarray}
where $F_\lambda=\partial F/\partial\lambda$, $F_\sigma=\partial F/\partial\sigma$ and $\xi_\sigma=\partial\xi/\partial\sigma$. If a combination $F_\lambda \xi_\sigma-F_\sigma^2$ does not vanish,
the two equations are independent and $\lambda$ and $\sigma$ are given
by
\begin{eqnarray}
&&\lambda=R, \\
&&\sigma=\GB.
\end{eqnarray}
Eliminating $\lambda$ and $\sigma$ in the original action by using
these results, we find $S$ in (\ref{action1}) reduces to
(\ref{action0}): the equivalence of these two actions also holds for the equations of
motion. If the combination $F_\lambda \xi_\sigma-F_\sigma^2$ vanishes,
(\ref{lambda1}) and (\ref{sigma1}) are not independent from each other.
This does not mean we can not eliminate $\lambda$ and $\sigma$ from
the original action.  Because (\ref{lambda1}) is the condition that
the action remains the same under the variation, all the possible
pairs of $\lambda$ and $\sigma$ that satisfy (\ref{lambda1}) give the
same action.  We can put any $(\lambda, \sigma)$ we want into $S$ as
long as $(\lambda, \sigma)$ are the solutions of (\ref{lambda1}).
Obviously, $\lambda=R$ and $\sigma=\GB$ are the solution and we can put
them into (\ref{action1}) to find that it recovers the original action
(\ref{action0}).

\subsection{Equations of motion}
The equations of motion for $g_{\mu \nu}$, in the presence also of a matter component with stress-energy tensor $T_{\mu\nu}$, are given by
\begin{eqnarray}
R_{\mu \nu}-\tfrac{1}{2} g_{\mu \nu}R-\Sigma_{\mu \nu}=
\frac{8\pi}{M_P^2\,F}\,T_{\mu\nu}, \label{eom}
\end{eqnarray}
where $\Sigma_{\mu \nu}$ is the effective energy momentum tensor defined by
\begin{eqnarray} 
\Sigma_{\mu \nu}&=&\frac1F \bigl(\nabla_\mu \nabla_\nu F-g_{\mu \nu} \Box F+2R \nabla_\mu \nabla_\nu \xi-2g_{\mu \nu} R \Box \xi-4R_\mu^{~\lambda} \nabla_\lambda \nabla_\nu \xi \nonumber \\
&&-4R_\nu^{~\lambda} \nabla_\lambda \nabla_\mu \xi 
+4R_{\mu \nu} \Box \xi+4 g_{\mu \nu} R^{\alpha \beta} \nabla_\alpha \nabla_\beta \xi+4R_{\mu \alpha \beta \nu} \nabla^\alpha \nabla^\beta \xi-\tfrac{1}{2}\,g_{\mu \nu} V \bigr). \label{effective-energy-momentum}
\label{enemom}
\end{eqnarray}
The equations of motions for $\lambda$ and $\sigma$ are instead given by $\lambda=R$ and $\sigma=\GB$.

\subsection{Background dynamics}
We assume that the background spacetime is a flat FLRW universe whose metric is
given by
\begin{eqnarray}
\de s^2=-\de t^2+a^2(t)\, \de x_i\, \de x^i, \label{bg1}
\end{eqnarray}
where the indices $i,j,...$ are raised by $\delta_{ij}$.
Then the background components are
\begin{align}
G^0_0&=-3H^2,\\
G^i_j&=-\left( H^2+2 \frac{\ddot a}{a} \right) \delta^i_j,\\
\Sigma^0_0&=\frac{1}{F} \left[ 3H {\dot F}+12 H^3 {\dot \xi}-\tfrac12\,V \right],\\
\Sigma^i_j&=\frac{1}{F} \left[ 2H{\dot F}+8H \frac{\ddot a}{a} {\dot \xi}+4 H^2 {\ddot \xi}+{\ddot F}-\tfrac12\,V \right] \delta^i_j,\\
R&=6(2H^2+{\dot H}),\\
\GB&=24 H^2 (H^2+{\dot H}), 
\end{align}
from which we have, in vacuum
\begin{align}
3H^2&=\frac{1}{F} \left[\tfrac12\, V-3H{\dot F}-12H^3 {\dot \xi} \right],\\
{\dot H}&=\frac{1}{2F+8H{\dot \xi}} \left[ -{\ddot F}+H{\dot F}-4H^2 ({\ddot \xi}-H{\dot \xi}) \right],\\
\lambda&=6(2H^2+{\dot H}),\\
\sigma&=24 H^2 (H^2+{\dot H}).
\end{align}

\section{Scalar Perturbation}

\subsection{Perturbed metric}
We consider scalar perturbations around the metric Eq.~(\ref{bg1}).
We write the perturbed metric as
\begin{eqnarray}
\de s^2=-(1+2\alpha)\, \de t^2-2 a (t) \partial_i \beta\, \de t\, \de x^i+a^2(t) (\delta_{ij}+2\phi \delta_{ij}+2\partial_i \partial_j \gamma)\, \de x^i \,\de x^j.\label{per1}
\end{eqnarray}
For later convenience, we define $\chi$ by the following equation,
\begin{eqnarray}
\chi \equiv a(\beta+a {\dot \gamma}).
\end{eqnarray}
This represents the shear potential of the unit vector normal to $\Sigma_t$, where $\Sigma_t$ is a time-like hypersurface of constant $t$.

\subsection{Gauge transformation}
There are degrees of freedom of choosing $\Sigma_t$. 
Changing from $\Sigma_t$ to ${\tilde \Sigma_t}$ corresponds to a time coordinate transformation: $t\to t+T(t,x^i)$. Under this transformation, the perturbation variables transform as
\begin{eqnarray}
&&{\tilde \alpha}=\alpha-{\dot T}, \\
&&{\tilde \phi}=\phi-H T, \\
&&{\tilde \chi}=\chi-T, \label{gtchi} \\
&&{\tilde {\delta F}}=\delta F-{\dot F}T, \\
&&{\tilde {\delta \xi}}=\delta \xi-{\dot \xi}T .
\end{eqnarray}

\subsection{Perturbation equations in general gauge}
The perturbations of the Einstein tensor can be written in Fourier space as
\begin{eqnarray}
&&\delta G^0_0=2 \left( 3H^2 \alpha-\frac{k^2}{a^2} \phi-3H{\dot \phi}+\frac{k^2}{a^2} H\chi \right),\label{pe1}\\
&&\delta G^0_i=2 \partial_i ( {\dot \phi}-H\alpha ), \label{pe2}\\
&&\delta G^i_j-\frac{1}{3}\delta^i_j \delta G^\ell_\ell=a^{-2} \left( \partial^i \partial_j-\frac{1}{3} \delta^i_j \triangle \right) ({\dot \chi}+H \chi-\phi-\alpha). \label{pe3}
\end{eqnarray}
It can be shown that under the gauge transformation $t\to t+T(t,x^i)$ the Einstein tensor transforms as
\begin{eqnarray}
&&{\tilde {\delta G^0_0}} = \delta G^0_0+6H {\dot H} T,\\
&&{\tilde {\delta G^0_i}} = \delta G^0_i-2 {\dot H} \partial_i T, \\
&&{\tilde {\delta G^i_j}}-\frac{1}{3} \delta^i_j {\tilde \delta G^\ell_\ell}=\delta G^i_j-\frac{1}{3}\delta^i_j \delta G^\ell_\ell.
\end{eqnarray}
We do not consider the trace part of $G^i_j$ because the perturbation
equations for the trace part can be derived by the combinations of
other equations and do not bring new informations.  We find that the
traceless part of $\delta G^i_j$ is gauge invariant.

The perturbations of $\Sigma^\mu_\nu$ given in Eq.~(\ref{effective-energy-momentum}) 
can be written as
\begin{align}
\delta\Sigma^0_0&=\frac{1}{F} \bigg[ 3({\dot F} +12H^2 {\dot \xi}) {\dot \phi}+8 \frac{k^2}{a^2} H {\dot \xi} \phi-6H ({\dot F}+8 H^2 {\dot \xi}) \alpha- \frac{k^2}{a^2}({\dot F}+12H^2 {\dot \xi}) \chi \nonumber \\
&\qquad+3H ({\dot {\delta F}}+4H^2 {\dot {\delta \xi}})+\left( -3 (H^2+{\dot H})+\frac{k^2}{a^2} \right) (\delta F+4H^2 \delta \xi) \bigg],\\
\delta\Sigma^0_i &=\frac{1}{F} \partial_i [-{\dot {\delta F}}+H\delta F-4H^2 {\dot {\delta \xi}}+4H^3 \delta \xi+({\dot F}+12H^2 {\dot \xi}) \alpha-8H {\dot \xi} {\dot \phi} ]\,,\\
\delta\Sigma^i_j-\frac{1}{3}\delta^i_j \delta \Sigma^\ell_\ell&=\frac{1}{a^2 F} \left( \partial^i \partial_j-\frac{1}{3} \delta^i_j \triangle \right) \bigg\{ 4H{\dot \xi}\alpha+4{\ddot \xi} \phi-[{\dot F}+4(H^2+{\dot H}){\dot \xi} +4H{\ddot \xi}]\,\chi \nonumber \\
&\qquad-4H{\dot \xi}{\dot \chi}+\delta F+4(H^2+{\dot H}) \delta \xi \bigg\}.
\end{align}
With the help of the background equations, we find that under the gauge transformation $t\to t+T(t,x^i)$, $\Sigma^\mu_\nu$ transform as
\begin{eqnarray}
&&{\tilde {\delta \Sigma^0_0}} = \delta \Sigma^0_0+6H {\dot H} T,\\
&&{\tilde {\delta \Sigma^0_i}} = \delta \Sigma^0_i-2 {\dot H} \partial_i T, \\
&&{\tilde {\delta \Sigma^i_j}}-\frac{1}{3} \delta^i_j {\tilde \delta \Sigma^\ell_\ell}
=\delta \Sigma^i_j-\frac{1}{3}\delta^i_j \delta \Sigma^\ell_\ell.
\end{eqnarray}
The explicit verification that both $G^\mu_\nu$ and $\Sigma^\mu_\nu$ transform in the same way under the gauge transformation supports that 
the derived perturbation equations are indeed correct.

Collecting these results, the perturbation equations, in Fourier space, are given by
\begin{align}
3H^2 \alpha&-\frac{k^2}{a^2} \phi-3H{\dot \phi}+\frac{k^2}{a^2} H\chi=\frac{1}{2F} \bigg[ 3({\dot F} +12H^2 {\dot \xi}) {\dot \phi}+8 \frac{k^2}{a^2} H {\dot \xi} \phi-6H ({\dot F}+8 H^2 {\dot \xi}) \alpha \nonumber \\
&\hspace{60mm}- \frac{k^2}{a^2}({\dot F}+12H^2 {\dot \xi}) \chi+3H ({\dot {\delta F}}+4H^2 {\dot {\delta \xi}}) \nonumber \\
&\hspace{60mm}+\left( -3 (H^2+{\dot H})+\frac{k^2}{a^2} \right) (\delta F+4H^2 \delta \xi) \bigg], \label{per00} \\
{\dot \phi}&-H\alpha =\frac{1}{2F} \bigg[H\delta F-{\dot {\delta F}}-4H^2 {\dot {\delta \xi}}+4H^3 \delta \xi+({\dot F}+12H^2 {\dot \xi}) \alpha-8H {\dot \xi} {\dot \phi} \bigg], \label{per0i}\\
{\dot \chi}&+H \chi-\phi-\alpha=\frac{1}{F} \bigg[ 4H{\dot \xi}\alpha+4{\ddot \xi} \phi-({\dot F}+4(H^2+{\dot H}){\dot \xi} +4H{\ddot \xi})\chi \nonumber \\
&\hspace{50mm}-4H{\dot \xi}{\dot \chi}+\delta F+4(H^2+{\dot H}) \delta \xi \bigg],\label{pertrls}\\
\delta \lambda&=-2 \left[ 6(2H^2+{\dot H}) \alpha+\frac{k^2}{a^2} ({\dot \chi}+2H\chi)-2\frac{k^2}{a^2} \phi-\frac{k^2}{a^2} \alpha-3{\ddot \phi}-12H{\dot \phi}+3H{\dot \alpha} \right], \label{perricci}\\
\delta \sigma&=-8 \bigg[ 12(H^2+{\dot H}) H^2 \alpha+2\frac{k^2}{a^2} H(H^2+{\dot H}) \chi+\frac{k^2}{a^2} H^2 {\dot \chi}-3H^2 {\ddot \phi}+3H^3 {\dot \alpha} \nonumber \\
&\hspace{40mm} -6H(2H^2+{\dot H}) {\dot \phi}-2\frac{k^2}{a^2} (H^2+{\dot H}) \phi-\frac{k^2}{a^2} H^2 \alpha \bigg]. \label{pergb}
\end{align}

\subsection{Gauge invariant variables and the master equation}

It is convenient to analyze the perturbation equations in terms of gauge invariant variables. Therefore, let us define the following gauge independent combinations of fields
\begin{eqnarray}
  \delta\Gamma &=& \dot F\,\delta\xi-\dot\xi\,\delta F\, ,\label{defgamma} \\
  \Phi &=& \phi-\frac{H (\delta F+4 H^2\delta\xi)}{\dot F+4 H^2 \dot\xi}\, ,\label{defphi} \\
  \Psi &=& \phi+\frac{\delta F+4 H^2\delta\xi}{2 (F+4 H \dot\xi)}
  +\tfrac{1}{2}\, \chi
  \left[\frac{F H-\dot F}{F+4 H \dot\xi}-3 H\right] \, . \label{defpsi}
\end{eqnarray}
We will see that these three fields are enough to understand the
behaviour of the scalar perturbations in the most general theory of
gravity $f(R,\GB)$. This field definition cannot be applied on
maximally symmetric backgrounds, where $H=H_0$, and $\dot
F=\dot\xi=0$. These backgrounds must be studied as a special case,
which will be done later.  In general, one can solve equation
(\ref{per0i}) for $\alpha$ and substitute it into equation
(\ref{per00}). Afterwards one can use the fields introduced before, to
change (\ref{per00}) into
\begin{equation}
  \label{eq:per00BB}
  \frac{k^2}{a^2}\,\Psi=A_1(t)\, \dot\Phi+A_2(t)\,\delta\Gamma\, ,
\end{equation}
where $A_{1,2}$ are functions of the background only and are defined in the appendix. It must be noticed that, in order to write down this equation, we made use of the equations of motion in order to replace $\ddot F$ in terms of the other backgrounds variables. Along the same lines one can rewrite equation (\ref{pertrls}) into the following form
\begin{equation}
  \label{eq:trlsBB}
  \dot\Psi=A_3(t)\,\delta\Gamma+A_4(t)\,\Psi+A_5(t)\,\Phi\, .
\end{equation}
We need at least another equation in order to make the equations
closed.  For this aim, we use then the definition of the field
$\delta\Gamma$ and write the following formula
\begin{equation}
  \label{eq:dG1}
  \delta\Gamma=(\dot F\xi_\sigma-\dot\xi F_\sigma)\delta\sigma+
  (\dot F\xi_\lambda-\dot\xi F_\lambda)\delta\lambda
=(F_\lambda\,\xi_\sigma-F_\sigma^2)\,(\dot\lambda\delta\sigma-\dot\sigma\delta\lambda)\, .
\end{equation}
It is possible to rewrite both $F_\lambda$ and $\xi_\sigma$ in terms of $\dot F$, $\dot \xi$, and $F_\sigma$ as follows
\begin{align}
  F_\lambda&=\frac{\dot F-F_\sigma\dot\sigma}{\dot\lambda}\, ,\\
  \xi_\sigma&=\frac{\dot\xi-F_\sigma\dot\lambda}{\dot\sigma}\, .
\end{align}
Therefore we can put equations (\ref{perricci}) and (\ref{pergb}) into
equation (\ref{eq:dG1}) to write another equation for $\delta\Gamma$
and the other fields. The equation that one obtains has the following
form
\begin{equation}
  \label{eq:dG2}
  p_1(t) \,\dot{\delta\Gamma}+p_2(t)\,\dot\Phi+p_3(t)\,\delta\Gamma
  +\frac{k^2}{a^2}\,[p_4(t)\,\delta\Gamma+\dot\Psi+p_5(t)\,\Psi+p_6(t)\,\Phi]
  +p_7(t)\,\ddot\Phi=0\, ,
\end{equation}
where the $p_i$ are all functions of time only. This complicated
equation can be simplified as follows. From equation (\ref{eq:trlsBB})
one can solve for $\dot\Psi=\dot\Psi(\Psi,\Phi,\delta\Gamma)$. 
By using equation (\ref{eq:per00BB}) one can find
$\dot\Phi=\dot\Phi(\Psi,\delta\Gamma)$. By differentiating once this
last equation and by replacing $\dot\Psi$ and $\ddot F$
(this last one, by using the background equations) one can also find
$\ddot\Phi=\ddot\Phi(\Phi,\Psi,\delta\Gamma,\dot{\delta\Gamma})$. 
Inserting this last equation into (\ref{eq:dG2}) to eliminate $\ddot \Phi$ and 
also replacing $\dot\Phi$ and $\dot\Psi$ with expressions that do not
contain time derivatives of the fields,
we have an equation given by
\begin{equation}
  \label{eq:dG3}
  \delta\Gamma=\frac{k^2}{a^2}\,[A_6(t)\,\Phi+A_7(t)\,\Psi]\, ,
\end{equation}
where $A_{6,7}$ are given in the appendix. We find that ${\dot {\delta \Gamma}}$
automatically disappears from the equation, which is crucial to
derive the closed second order differential equation.  As we will see
later on, some gravity models identically yield $\delta \Gamma=0$.  In
these models, we can verify that $A_6$ and $A_7$ also identically
vanish and (\ref{eq:dG3}) does not give any information.

Now, it is possible to find a closed second order differential
equation in time for the Fourier fields $\Phi$, or $\Psi$, or $\delta\Gamma$. For
example, let us use equation (\ref{eq:dG3}) into equations
(\ref{eq:trlsBB}) and (\ref{eq:per00BB}) obtaining
\begin{align}
  \label{eq:BB2}
    \frac{k^2}{a^2}\,[1-A_2(t) A_7(t)]\Psi&=A_1(t)\, \dot\Phi+A_2(t)\,A_6(t)\,\frac{k^2}{a^2}\,\Phi\, ,\\
    \label{eq:TR2}
    \dot\Psi&=\frac{k^2}{a^2}\,A_3(t)\,A_6(t)\,\Phi+\frac{k^2}{a^2}\,A_3(t)\,A_7(t)\,\Psi+A_4(t)\,\Psi+A_5(t)\,\Phi\, .
\end{align}
Then let us solve (\ref{eq:BB2}) for $\Psi$, finding $\Psi=\Psi(\Phi,\dot\Phi)$. Then we can use this result into equation (\ref{eq:TR2}) to find an equation of the form
\begin{equation}
  \label{eq:ScGen}
  \frac1{a^3\, Q(t)}\partial_t[a^3\, Q(t)\,\dot\Phi]+B_1(t)\, \frac{k^2}{a^2}\,\Phi+B_2(t)\,\frac{k^4}{a^4}\,\Phi=0\, , 
\end{equation}
where $Q$, $B_1$, and $B_2$ are defined in the appendix.  The field
$\Phi$ indeed seems to be a ``good'' field to study in the sense that
its equation of motion has a relatively simple dependence in time and
$k$.

Eq. (\ref{eq:ScGen}) is the main result of this paper.  The scalar
perturbations have only two independent degrees of freedom, which are
$\Phi(t_0)$ and $\dot\Phi(t_0)$.  All the informations of the
perturbation behavior are completely determined by (\ref{eq:ScGen}).
To see this, looking at equation (\ref{eq:BB2}) and (\ref{eq:dG3}), 
one can write both $\Psi$ and $\delta \Gamma$ as linear combinations of $\Phi$ and ${\dot \Phi}$.  As for the metric
perturbation variables themselves such as $\alpha, \chi$ and $\phi$,
because of their gauge dependence, they cannot be written in terms of
$\Phi, \Psi$ and $\delta \Gamma$ which are gauge invariant.  But this
does not mean that we need further knowledges which are not contained
in $\Phi, \Psi$ and $\delta \Gamma$.  It merely means that only the
gauge invariant combinations out of $\{ \alpha, \chi, \phi \}$, or
equivalently $\{ \alpha, \chi, \phi \}$ in a specific gauge, are
expressed by $\Phi, \Psi$ and $\delta \Gamma$.  As an illustration,
let us consider a gauge where $\delta F=-4H^2 \delta \xi$, which we
call the MGM Gauge (MGMG).  There are no remaining
gauge degrees of freedom in this gauge.  In this gauge, we have
$\phi=\Phi$ from (\ref{defphi}).  By using the remaining two equations
(\ref{defgamma}) and (\ref{defpsi}), $\delta \xi$ and $\chi$ are
uniquely expressed in terms of $\Phi, \Psi$ and $\delta \Gamma$.
After that, $\alpha$ is given by (\ref{pertrls}) and finally all the
metric perturbation variables are determined.  
Once the metric perturbations are determined in a specific gauge,
those in other gauges are simply obtained by gauge transformation.

Therefore, the knowledge of $\Phi$ and ${\dot \Phi}$ is enough to understand the
behavior of the metric perturbations.

The master equation (\ref{eq:ScGen}) contains a term proportional to $k^4$,
or equivalently a term of fourth order spatial derivative in real
space.  This term does not vanish in generic $f(R,\GB)$ models, nor is a
spurious result due to a bad choice of fields/gauge: the field $\Phi$
is gauge invariant, and this $k^4$ behaviour would still be there by
studying either $\delta\Gamma$ or $\Psi$. Furthermore in the MGMG
$\Phi$ has a simple meaning which can be directly related to
experimental data.  Therefore, for the general theory, we don't have a
standard wave equation and we expect that the perturbations propagate
in space in non-trivial ways.  

It should be noted, however, that $B_2 (t)$ identically vanishes
if $F_\lambda \xi_\sigma-F_\sigma^2=0$.  Interestingly, most of the
modified gravity models concerned by the past literature, such as
$f(R)$ gravity and $R+f(\GB)$ gravity models, belong to these special
cases.  In these special cases, the propagation properties of the
perturbations deeply differ from those of the generic cases.  
We defer to study the special cases in section VI.  
In this section, we consider the generic cases where $B_2(t) \neq 0$.

The extreme complexity of $B_1(t)$ and $B_2(t)$ requires numerical
calculations to solve exactly the differential equation and to see how the
perturbations evolve.  However, studies for limiting cases still allow
us to obtain important informations such as instability of the
perturbations, which will be done in section IV.

\subsection{Action of the perturbations}

By expanding the action (\ref{action1}) up to second order in the perturbation
variables and eliminating all the auxiliary fields using the equations of motion,
we find that the action in terms of $\Phi$ can be written as 
\begin{equation}
  \label{eq:actPhi}
  S=\frac{M_P^2}{\pi}\int dt\,d^3x\,Q\,a^3\left[\frac12\,\dot\Phi^2
    -\frac12\,\frac{B_1}{a^2}(\vec\nabla\Phi)^2-\frac12\,\frac{B_2}{a^4}(\vec\nabla^2\Phi)^2\right]\, .
\end{equation}
From this action, we correctly recover the master equation
(\ref{eq:ScGen}).  We find that the sign of the kinetic term in the
action is equal to that of $Q$.  We will then call that mode for which
its kinetic energy has a negative sign, that is $Q<0$, a ghost.
Typically, if a ghost is present and coupled with other normal fields,
we can expect that the vacuum decays into metric and normal fields,
because the energy conservation allows such a process to happen.
Therefore, one would expect strong bounds from the vacuum decay
\cite{Cline:2003gs}.  Here we give the condition $Q>0$ as a necessary
one in order to have a theory without ghosts degrees of
freedom. However, we are not interested in the details of the
quantization procedure which is outlined in
\cite{Hwang:2005hb}. Although the quantization mentioned in
\cite{Hwang:2005hb} only treats the case $B_2=0$, the $k^4$ term will
only change the dispersion relation $\omega_k(t)$ for each decoupled
mode. Therefore, this change will only lead to an explicit expression
for the solution $\bar\Phi_k(t)$ which is, in general, different from
the $B_2(t)=0$ case.  Anyhow, we will leave this point in detail for a
future project, when matter fields will be introduced. In summary, the
first bound we will consider is the one coming from not having ghosts
in the theory, that is requiring a positive $Q$.

\section{Study of The Master Equation}
\subsection{Long wavelength limit}
Let us first study a case where the wavelength of the mode we consider
is much larger than any typical length scale.  In this case,
neglecting terms that are quadratic and quartic in $k/a$ in
(\ref{eq:ScGen}) yields
\begin{eqnarray}
\frac1{a^3\, Q(t)}\partial_t[a^3\, Q(t)\,\dot\Phi]=0.
\end{eqnarray}
We can immediately integrate this differential equation and the result is
\begin{eqnarray}
\Phi=C_k+D_k \int_0^t \frac{dt'}{a^3 (t') Q(t')},
\end{eqnarray}
where $C_k,D_k$ are independent of $t$, being integration constants for each $k$.

$\Psi$ can be determined by substituting the above solution into (\ref{eq:BB2}).
To the leading order,
$\Psi$ is given by
\begin{eqnarray}
\Psi \approx \frac{1}{1-A_2 A_7} \left( A_1 \frac{D_k}{a^3 Q} \frac{a^2}{k^2}+A_2 A_6 C_k \right).
\end{eqnarray}
Depending on the leading power of $k$ for both $C_k$ and $D_k$, each term on the right hand side can be dominant.

\subsection{Short wavelength limit}
Let us next study a case where the wavelength of the mode we consider
is much smaller than any typical length scale.  In the large $k$ limit,
the time scale for the change of $\Phi$ is much smaller than that for
the change of the background quantities.  Therefore, we can use WKB
approximation to obtain the solutions. The solutions of
(\ref{eq:ScGen}), under the WKB approximation, can be given by
\begin{eqnarray}
\Psi (t) &\approx& c_+ \exp \bigg[ i \int^t dt'~\left( \omega_k^{(+)}+\frac{i {\dot \omega_k^{(+)}}}{2\omega_k^{(+)}-i {(a^3 Q)}^\cdot /(a^3 Q)} \right) \bigg] \nonumber \\
&&+c_- \exp \bigg[ -i \int^t dt'~\left( \omega_k^{(-)}-\frac{i {\dot \omega_k^{(-)}}}{2\omega_k^{(-)} -i{(a^3 Q)}^\cdot /(a^3 Q)} \right) \bigg],
\end{eqnarray}
where $c_\pm$ are constants and $\omega_k^{(\pm)}$ are the roots of the following quadratic equation
\begin{eqnarray}
\omega^2-i \frac{ {(a^3 Q)}^\cdot }{a^3 Q}\omega -\frac{k^4}{a^4} B_2-\frac{k^2}{a^2} B_1=0.
\end{eqnarray}
In the large $k$ limit, $\omega_k^{(\pm)}$ are given by
\begin{equation}
\label{eq2ndD}
\omega_k^{(\pm)}=\pm \sqrt{B_2}\, \frac{k^2}{a^2} \pm \frac{B_1}{2 \sqrt{B_2}}+\frac{i}{2}\frac{ {(a^3 Q)}^\cdot }{a^3 Q} +{\cal O}(k^{-2}).
\end{equation}

We see that if $B_2$ is negative, then $\omega_k^{(\pm)}$ are pure
imaginary.  Therefore, $\Phi$ grows exponentially in time.  The growth
rate increases in proportion to $k^2$.  The smaller the wavelength of
the mode is, the smaller the time scale of the instability is.  One
may guess this instability is due to a bad choice of perturbation
variables and there might exist a special gauge where all the metric
perturbation variables do not exhibit such an instability, i.e.\ where
the deviations from the FLRW universe remain small.  If this were
possible, $\Phi$, written in terms of the metric perturbation
variables in that gauge, should not show an exponential growth, which
is inconsistent with the exponential growth of $\Phi$ we have just
found.  Therefore this instability is not a due to a bad choice for
the perturbation variables and the FLRW universe is indeed unstable on
small scales.  Due to this instability, the perturbations will grow
until they become ${\cal O}(1)$ where the linear perturbation theory
no longer works.  We do not know what happens if the system goes into
the non-linear regime and will not consider it furthermore.

If $B_2$ is positive, then the leading term of $\omega_k^{(\pm)}$ is
real and the perturbations propagate.  In this case, the group velocity is given by
\begin{eqnarray}
v_g (k)= a \frac{\partial | \omega_k^{(\pm)} |}{\partial k} \approx 2 \sqrt{B_2}\, \frac{k}{a}.
\end{eqnarray}
At any time $t_0$, this velocity exceeds the speed of light for those
modes above a critical wavenumber $k_c$ which is given by
$k_c=a(t_0)/[2 \sqrt{B_2(t_0)}]$.  The propagation of the short
wavelength modes inevitably becomes superluminal up to the cutoff of
the theory $v_{g,{\rm max}}<2\sqrt{B_2}\Lambda_{\rm cutoff}$.  The
smaller the wavelength of the mode is, the larger the propagation
speed. For a given $k$, the speed of propagation is background
dependent. Because of this, if $v_g$ reduces quickly with time, for some backgrounds, it may be possible that the friction term (the third one in the right hand side of Eq.\ \ref{eq2ndD}) gives the strongest contribution to the stability of each mode, at least for $t\to\infty$. It still remains a lack of consensus among researchers if
the propagation speed larger than the speed of light causes crucial
problems that immediately make us give up considering such models as a
possible modification for a consistent theory of gravity. More details are given in the discussion section.

We will also impose $B_1$ to be positive as there is always some
intermediate range of $k$'s, for which the $B_1k^2$ term will be
dominating over the $B_2k^4$ one.

To summarize, for the general $f(R,\GB)$ MGM models, except for those
special cases where $F_\lambda \xi_\sigma-F_\sigma^2 = 0$, the small
wavelength modes inevitably either suffer from strong instability or
acquire superluminal propagation.

\section{Application to a toy model}

Just for the purpose of illustrating how the general formalism we have
developed in the previous section can be applied to an explicit MGM,
let us consider the following toy model, introduced in (\cite{Carroll:2004de}), whose $f(R,\GB)$ is given by,
\begin{eqnarray}
f(R,\GB)=R+\frac{M^6}{R^2+\alpha \GB}, \label{toymodel}
\end{eqnarray}
where $M$ is a constant of mass dimension and $\alpha$ is a
dimensionless constant.  In this model, we have that
\begin{eqnarray}
\frac{\partial^2 f}{\partial R^2} \frac{\partial^2 f}{\partial \GB^2}-{\left( \frac{\partial^2 f}{\partial R \partial \GB} \right)}^2=-\frac{4M^{12} \alpha^2}{ {(R^2+\alpha \GB)}^5},
\end{eqnarray}
which does not vanish in general. Therefore, this model does not
belong to the special class of MGM, and its master equation will have the $k^4$-term.

It is known that the background equations can have power-law solution
such as $a(t) \propto t^p$ for four different values of $p$.  They are
given by
\begin{align}
p_1&=\frac{51+10 \alpha+y(\alpha)}{4(6+\alpha)}, \\
p_2&=\frac{51+10 \alpha-y(\alpha)}{4(6+\alpha)}, \\
p_3&=\frac{6+\alpha+\sqrt{6\alpha+\alpha^2}}{2 (6+\alpha)}, \\
p_4&=\frac{6+\alpha-\sqrt{6\alpha+\alpha^2}}{2 (6+\alpha)}, \label{p4}
\end{align}
where $y=y(\alpha)=\sqrt{1521+840 \alpha+100 \alpha^2}$. If the radicand
becomes negative, which indeed happens for some values of $\alpha$, the
corresponding power-law solution will not exist.

From now on, we only consider the case $p=p_1$.  Analysis for other
$p_i$ can be done in exactly the same way.  From the background
equation, we find that the power-law solution is an attractor if the
following inequality
\begin{equation}
\frac{273+50\alpha+3 y(\alpha)}{6+\alpha}>0, \label{attractor}
\end{equation}
is satisfied.  If not, the universe cannot settle down to this
power-law solution. From the definition of $y(\alpha)$, we find that
$y(\alpha)$ is real for $\alpha \le -3(14+3\sqrt{3})/10 \approx -5.76$
and for $\alpha \ge 3(-14+3\sqrt{3})/10 \approx -2.64$.  For such
ranges of $\alpha$, the power-law solution becomes an attractor for
$\alpha \le -507/80 \approx -6.34$ and for $\alpha \ge -2.64$.
Moreover, such an attractor is an accelerating one for $\alpha \ge
-2.64$, whereas it is a decelerating one for $\alpha \le -6.34$.

\begin{figure}[t]
\begin{center}
\includegraphics[width=10cm,keepaspectratio]{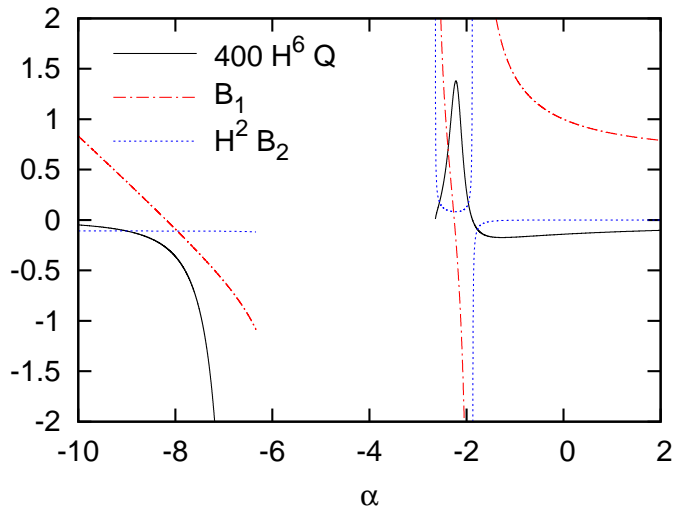}
\caption{Plot of $Q,B_1,B_2$ for one attractor of the toy model
  considered. $B_1$ becomes positive for three regions of $\alpha$:
  (i) $\alpha < -8.19$, (ii)$-2.64 < \alpha < -2.25$, (iii) $\alpha >
  -1.875$. $H^2 B_2$ becomes positive only for $-2.64 < \alpha <
  -1.875$. $H^6Q$ becomes positive only for $-2.64 < \alpha < -1.875$.}
\label{graphB1}
\end{center}
\end{figure}

Using the formulae given in the appendix, the expressions for $B_1$,
$B_2$ and $Q$ at late time $(t \to \infty)$ are given by
\begin{align}
B_1&=\frac{135(26364-391 y)+16\alpha \{60 (3660-49 y)+\alpha [70716-631 y+10 \alpha (907+40\alpha-4 y)]\}}{45(15+8\alpha)(57+8\alpha) y}, \\
H^2B_2&=\frac{-16 \alpha^2 (6+\alpha) (51+10\alpha+y)}{3 [8\alpha (15+2\alpha)+3 (39+y)] \{351(39+y)+2\alpha [5301+87 y+10\alpha (129+10 \alpha+y)]\}}, \\
H^6Q &=\frac{10125 (15+8\alpha) (57+8\alpha) y}{4s(\alpha)}, \\
s(\alpha)&=-4108137345 (39+y)-2 \alpha \bigl\{-9477 (-18610644+44279 y)+10 \alpha \bigl[-297 (-46531983+545000y) \nonumber \\
&\quad+320 \alpha \bigl(27 (579515-6608y)+5\alpha [581040-4563 y+10 \alpha (5403+200\alpha-20 y)]\bigr)\bigr]\bigr\}.
\end{align}
Here we have multiplied $B_2$ and $Q$ by $H^2$ and $H^6$ to make them
dimensionless.  Graphs of these quantities as functions of $\alpha$
are shown in Fig.\ \ref{graphB1}.

Here we find that the region of $\alpha$ where all of the quantities
become positive is $-2.64 < \alpha < -2.25$. This range is outside the
allowed range coming from different constraints regarding the fit to
Supernovae Type Ia data and the viability of the cosmological
evolution from radiation domination up to today
\cite{Mena:2005ta,DeFelice:2007zq}.  Therefore, according to our
results, for these models, in the $\alpha$-interval allowed by
theory/experiment, the background attractor $p_1$ will be in general
unstable. In fact, outside this region, at least one of $\{ B_1, B_2,
Q \}$ becomes negative. The asymptotic value of $H^2 B_2$ for $\alpha
\to -\infty$ is $-1/9$.  At first glance, this is incompatible because
the action (\ref{toymodel}) in the limit $\alpha \to \pm \infty$
reduces to a form $f(R,\GB) \to R+M^6/(\alpha\, \GB)$ which belong to
the special case.  The origin of this gap comes from the naive
expectation that $R^2$ in the denominator of (\ref{toymodel}) will be
much smaller than $\alpha\, \GB$ for $| \alpha | \gg 1$ and can be
neglected.  This expectation is true for positively large $\alpha$.
For negatively large $\alpha$, $p_1$ becomes $9/(4\alpha)$ and $R^2$
in the denominator of (\ref{toymodel}) remains of the same order of
magnitude as $\alpha\, \GB$.  Therefore, as long as this this
background solution is concerned, the perturbation behaviors in the
limit $-\alpha \gg 1$ do not reduce to the special case.

\section{Special cases}

In the previous section, we found that the perturbations propagate in
non-trivial ways on small scales due to the $k^4$-term in the master
equation (\ref{eq:ScGen}).  However, there are special cases where
$B_2$, the coefficient in front of $k^4 \Phi$, vanishes identically.
In these cases, we do not have the exotic behavior such as a strong
instability or a superluminal propagation observed in the generic
models due to the $k^4$-term.  Instead of the $k^4$-term, the
$k^2$-term becomes important.  In this section, we study these special
cases.

From the explicit expression of $B_2$ given in the appendix, we see
that the special case happens if either ${\dot H}=0$ or $F_\lambda
\xi_\sigma-F_\sigma^2=0$.  Therefore, de Sitter space for any $f(R,\GB)$
MGM belongs to this case.  We can easily verify that both $R+f(\GB)$ and
$f(R)$ models for any background solutions also belong to this case.
As we will explain later, there are other infinite numbers of $f(R,\GB)$
models that, irrespective of the background solutions, yield $B_2=0$.
We will provide a systematic method to find those special gravity
models and will list some explicit forms of $f(R,\GB)$.

Let us comment that vanishing $B_2$ is equivalent to $\delta
\Gamma=0$.  To see this, let us first assume $B_2=0$.  This implies
that ${\dot H}=0$ or $F_\lambda \xi_\sigma-F_\sigma^2=0$.  If ${\dot
  H}=0$, then ${\dot \lambda}$ and ${\dot \sigma}$ are also zero.
From (\ref{eq:dG1}), we immediately find that $\delta \Gamma$ is also
zero.  If $F_\lambda \xi_\sigma-F_\sigma^2=0$, (\ref{eq:dG1}) tells us
again that $\delta \Gamma=0$.  Therefore, $B_2=0$ implies $\delta
\Gamma=0$.  Next let us assume that $\delta \Gamma=0$.  This implies
that either of $F_\lambda \xi_\sigma-F_\sigma^2=0$ or ${\dot
  \lambda}\, \delta \sigma-{\dot \sigma}\, \delta \lambda=0$.  In the
first case, it is obvious from the expression of $B_2$ given in the
appendix that $B_2$ vanishes.  In the second case, both ${\dot
  \lambda}$ and ${\dot \sigma}$ must vanish, which implies ${\dot
  H}=0$.  This again gives $B_2=0$.  Therefore, $B_2=0$ is equivalent
to $\delta \Gamma=0$.

Since there are two independent conditions for the models to belong to this special case,
we will study them separately.

\subsection{Special cases I: ${\dot H}=0$}
If the background space-time is de Sitter, we have ${\dot
  \lambda}={\dot \sigma}=0$. Since $F$ and $\xi$ are both functions of
$\lambda$ and $\sigma$ only, we also have that ${\dot F}={\dot
  \xi}=0$. Except for the scale factor, which varies like $a(t)
\propto e^{Ht}$, all the other background quantities appearing in the
perturbed equations are constants. Keeping this in mind, the perturbed
equations (\ref{per00}), (\ref{per0i}), (\ref{pertrls}),
(\ref{perricci}) and (\ref{pergb}) become
\begin{align}
3H^2 \alpha&-\frac{k^2}{a^2} \phi-3H{\dot \phi}+\frac{k^2}{a^2} H\chi=\frac{1}{2F} \bigg[
3H ({\dot {\delta F}}+4H^2 {\dot {\delta \xi}})
+\left( -3 H^2+\frac{k^2}{a^2} \right) (\delta F+4H^2 \delta \xi) \bigg], \label{per00DS} \\
{\dot \phi}&-H\alpha =\frac{1}{2F} \bigl[H\delta F-{\dot {\delta F}}
-4H^2 {\dot {\delta \xi}}+4H^3 \delta \xi\bigr], \label{per0iDS}\\
{\dot \chi}&+H \chi-\phi-\alpha=\frac{1}{F} \bigl[\delta F
+4H^2 \delta \xi \bigr],\label{pertrlsDS}\\
\delta \lambda&=-2 \left[ 12H^2 \alpha+\frac{k^2}{a^2} ({\dot \chi}+2H\chi)-2\frac{k^2}{a^2} \phi
-\frac{k^2}{a^2} \alpha-3{\ddot \phi}-12H{\dot \phi}+3H{\dot \alpha} \right], \label{perricciDS}\\
\delta \sigma&=-8 \bigg[ 12H^4 \alpha+2\frac{k^2}{a^2} H^3 \chi+\frac{k^2}{a^2} H^2 {\dot \chi}
-3H^2 {\ddot \phi}+3H^3 {\dot \alpha}
-12H^3 {\dot \phi}-2\frac{k^2}{a^2} H^2 \phi-\frac{k^2}{a^2} H^2 \alpha \bigg]. \label{pergbDS}
\end{align}
These equations can be rewritten in terms of gauge independent fields. 
We can adopt the same field $\Psi$ as in the general case (but evaluated on the de Sitter background), 
however $\Phi$ is not well defined for a de Sitter background. 
We will choose then another field $\PhiB$ defined as
\begin{equation}
  \label{eq:phibar}
  \PhiB = -\frac{\delta F+4H^2\delta\xi}{2F}\, .
\end{equation}
This combination is gauge invariant on de Sitter and reduces to $\phi$ in the Newtonian gauge ($\chi=0$). Therefore the equations of motion can be rewritten as 
\begin{align}
  \Psi&=0\, ,\\
  \ddot\Phi_{\textsc{DS}}&=-3H\dot\Phi_{\textsc{DS}}-\left[ \frac{k^2}{a^2}-4H^2
   +\frac{F}{3(F_\lambda+8H^2 F_\sigma+16H^4 \xi_\sigma)} \right]\PhiB\, \label{master-dS},
\end{align}
where $\delta F$ and $\delta\xi$ are not independent from each other as, on this background, we have $\delta\sigma=4H^2\delta\lambda$. Eq.~(\ref{master-dS}) is equal to the Klein-Gordon equation in de Sitter 
space-time with an effective mass 
\begin{eqnarray}
m_{\mathrm{eff}}^2=-4H^2+\frac{F}{3(F_\lambda+8H^2 F_\sigma+16H^4 \xi_\sigma)}.
\end{eqnarray}
If $m_{\rm eff}^2$ is negative, then the long wavelength modes are
unstable (tachyonic instability).  For any $f(R,\GB)$ gravity models,
the short wavelength modes are stable and the sound velocity is unity. The last quantity we need to find for this special case, is the expression for $Q$ for $\PhiB$. Its action can be written as
\begin{align}
  S&=\frac{M_P^2}\pi\int dt\,d^3x\,Q\,a^3\left[\frac12\,\dot\Phi_{\textsc{DS}}^2
    -\frac12\,\frac{1}{a^2}\,(\vec\nabla\PhiB)^2
    -\frac12\,m_{\mathrm{eff}}^2\,\PhiB^2\right] ,
\end{align}
where
\begin{equation}
  \label{eq:Qds}
  Q=\tfrac34\,F\, .
\end{equation}
Therefore $F$ must be positive in order to avoid ghost-like degrees of freedom to propagate.

\subsection{Special cases II: $F_\lambda \xi_\sigma-F_\sigma^2=0$}

Before we study the perturbation equation, let us try to find what
kinds of models belong to this case.

Because $F$ and $\xi$ are functions of $\lambda$ and $\sigma$,
we have
\begin{align}
\delta F&=F_\lambda \delta \lambda+F_\sigma \delta \sigma, \\
\delta\xi&=F_\sigma \delta \lambda+\xi_\sigma \delta \sigma.
\end{align}
Here we have used the identity $\xi_\lambda=F_\sigma$ in the equation
for $d\xi$.  Because $F_\lambda \xi_\sigma-F_\sigma^2$ is the
determinant of the matrix of the linear transformation above, the
condition $F_\lambda \xi_\sigma-F_\sigma^2=0$ tells us that $\delta F$ and
$\delta\xi$ are not independent.  This implies that $F$ is a function of $\xi$.
In other words,
$F$ and $\xi$ are related to each other by a single variable $\varphi$
like $F=F(\varphi),~\xi=\xi(\varphi)$.
Such modified gravity models are equivalent to the action
\begin{eqnarray}
S=\frac{M_P^2}{16\pi} \int d^4x\sqrt{-g}\,[ {\tilde F} (\varphi) R+{\tilde \xi} (\varphi) \GB-{\tilde U} (\varphi)], \label{action-special}
\end{eqnarray}
where ${\tilde F},~{\tilde \xi}$ and ${\tilde U}$ are arbitrary functions.
To see that this action is equivalent to the $f(R,\GB)$ action
for the special case,
let us first take the variation of (\ref{action-special}) with respect to
$\varphi$.
After the variation,
we obtain an equation given by
\begin{eqnarray}
R \frac{\de {\tilde F}}{\de\varphi}+\GB \frac{\de {\tilde \xi}}{\de\varphi}
-\frac{\de {\tilde U}}{\de\varphi}=0. \label{alge-eq}
\end{eqnarray}
By solving this algebraic equation with respect to $\varphi$, we can
write $\varphi$ as a function of $R$ and $\GB$, i.e.\ $\varphi=\varphi
(R,\GB)$.  Substituting this solution into (\ref{action-special}), we
find that it reduces to the $f(R,\GB)$ action.  Then we can calculate $F$ and $\xi$ as
\begin{align}
F&=\frac{\partial f}{\partial R}={\tilde F}\bigl(\varphi(R,\GB)\bigr), \\
\xi&=\frac{\partial f}{\partial G}={\tilde \xi}\bigl(\varphi(R,\GB)\bigr).
\end{align}
Therefore, ${\tilde F}$ and ${\tilde \xi}$ are equal to $F$ and $\xi$.
From these equations, we can easily verify that this model belongs to
the special case, since
\begin{eqnarray}
\frac{\partial F}{\partial R}\frac{\partial \xi}{\partial \GB}-\frac{\partial F}{\partial \GB}\frac{\partial \xi}{\partial R}=0.
\end{eqnarray}

In this way, we can construct $f(R,\GB)$ models that belongs to the
special case.  Though finding analytic solution of (\ref{alge-eq}) is
difficult in general, we can in principle find as many models as we
want by doing the above procedures for various ${\tilde F},~{\tilde
  \xi}$ and ${\tilde U}$.  In this paper, we provide only a few simple
$f(R,\GB)$ models by using the above method.  The first example is a
case where ${\tilde \xi}(\varphi)=c$ (c is a dimensionless constant).
In this case, (\ref{alge-eq}) tells us that $\varphi$ depends only on
$R$.  Therefore, the corresponding modified gravity model is
$f(R,\GB)=f(R)+c\,\GB$.  Note that the term $c\,\GB$ only adds a total
derivative in the action and does not contribute to the equation of
motion.  Therefore, this model is equivalent to the $f(R)$ gravity
model. The second example is the opposite of the first case, i.e
${\tilde F}(\varphi)=c$.  The corresponding modified gravity model is
$f(R,\GB)=c\,R+f(\GB)$.  The third example is a case where ${\tilde
  F}(\varphi)=M^2 {\tilde \xi}(\varphi)$ (M is a constant of mass
dimension).  In this case, (\ref{alge-eq}) tells us that $\varphi$
depends only on the combination $R+\GB/M^2$.  The corresponding modified
gravity model is $f(R,\GB)=f(R+\GB/M^2)$.  We list these models in
Table~\ref{table}. For the special models, the perturbation equation (\ref{eq:ScGen}) can now be written as
\begin{eqnarray}
\frac1{a^3\, Q(t)}\partial_t[a^3\, Q(t)\,\dot\Phi]+B_1(t) \frac{k^2}{a^2} \Phi=0,
\end{eqnarray}
where $B_1$ in these cases is significantly reduced to a much simpler expression
than the general case and is given by
\begin{eqnarray}
B_1 (t)=\frac{16 H {\dot \xi}[{\dot F}+4(H^2+{\dot H}) {\dot \xi}]+F[3{\dot F}+4(3H^2+4{\dot H}){\dot \xi}]-4({\dot F}+4H^2 {\dot \xi}){\ddot \xi}}{3(F+4H {\dot \xi})({\dot F}+4H^2 {\dot \xi})}. \label{b1}
\end{eqnarray}
In particular, for $f(R)$ gravity models, it reduces to
\begin{eqnarray}
B_1(t)=1.
\end{eqnarray}
Therefore, we correctly recover the well-known result that the
propagation speed is equal to the velocity of light in $f(R)$ models.

For $c\,R+f(\GB)$ gravity models, the propagation speed was calculated in
\cite{Hwang:2005hb}. In this case, according to our formula (\ref{b1}), $B_1 (t)$ reduces to
\begin{eqnarray}
B_1(t)=1+\frac{2 {\dot H}}{H^2}\, .
\end{eqnarray}
If the universe, in vacuum, accelerates more slowly than $a(t) \propto
t^2$, $B_1$ becomes negative and the FLRW universe is unstable on
small scales. Therefore for these theories, acceleration (faster than
$t^2$) is a condition for vacuum stability. On the other hand, if the
universe undergoes super-acceleration ${\dot H}>0$, the propagation of
the perturbations becomes superluminal.

For $f(R+\GB/M^2)$ gravity models, $B_1(t)$ reduces to
\begin{eqnarray}
B_1(t)=1+\frac{8{\dot H}}{M^2+4H^2}.
\end{eqnarray}
Therefore, similarly to the $c\,R+f(\GB)$ models, the propagation becomes superluminal if the universe undergoes super-acceleration.

\begin{table}[t]
\begin{center}
\begin{tabular} {|c|c|}\hline
Special $f(R,\GB)$ models & remark  \\ \hline
$f(R)+\GB /M^2 $ & ${\tilde \xi}(\varphi)=1/M^2$,~~~ $M$:Constant of mass dimension \\ \hline
$c\,R+f(\GB)$ &  ${\tilde F}(\varphi)=c$, ~~~ $c$:Dimensionless constant\\ \hline
$f(R+\GB/M^2)$ & ${\tilde F}(\varphi)=M^2 {\tilde \xi}(\varphi)$ \\ \hline
\end{tabular}
\caption{We list some of the special $f(R,\GB)$ models for which $B_2 (t)$ identically vanishes.  
  In the first case, different values of $M$ give the same
  modified gravity model because a linear term in $\GB$ in the action is
  a total derivative and does not contribute to the equations of
  motion.}
\label{table}
\end{center}
\end{table}


\section{Discussion}

So far we have considered the scalar modes. In the literature
\cite{Cartier:2001is,Hwang:2005hb}, it has been claimed that the
cosmological scalar perturbations for a general Lagrangian of the kind
$f(\phi,R)+\xi(\phi)\,\GB-U(\phi)$, were studied\footnote{Eliminating
  the auxiliary field $\phi$ by using the equation of motion, we find
  that this model is equivalent to the general $f(R,\GB)$ gravity
  models.}.  However, we think that the scalar perturbations of this
Lagrangian were not studied there. In fact, the authors considered
only the following Lagrangian $F(\phi)\,R+\xi(\phi)\,\GB-U(\phi)$,
which actually reduces to the special case discussed above. By doing
so, in the literature, the $k^4$-term has always been neglected, and
the most general MGM was never then fully studied. In fact, the
novelty of our analysis resides in the study of the general case where
the modes do get a non-trivial modification for the dispersion
relation.

As for the vector modes, they are not important since, as already
stated in \cite{Hwang:1999gf}, they do not propagate. We have
explicitly verified that this remains true even for the general cases
for which $F_\lambda\xi_\sigma-F_\sigma^2\neq0$ and $\dot H\neq0$. The
tensor modes instead, do propagate, but, differently from the scalars,
we checked that there is no difference between general and special
cases, and that their evolution equation coincides with the one given
in \cite{Hwang:2005hb}. More in detail their equations are
\begin{equation}
  \label{eq:TT1}
  \frac1{a^3Q_{TT}}\,\partial_t(a^3Q_{TT}\dot C_{\mu\nu})-\frac{c_{TT}^2}{a^2}\,\vec\nabla^2C_{\mu\nu}=0 ,
\end{equation}
where
\begin{align}
  Q_{TT}&=\tfrac12\,F+2H\dot\xi\, ,\\
  c_{TT}^2&=\frac{F+4\ddot\xi}{F+4H\dot\xi}\, .
\end{align}
However, they still give an important contribution in order to set
constraints for these theories. To summarize, we can then set the
following points in order to understand the vacuum structures for the
MGM.
\begin{itemize}
\item In the general case one must set the following conditions for
  the tensor modes $Q_{TT}>0$, $c_{TT}^2>0$. The scalar modes instead
  have a more complicated structure as they will acquire a non-trivial
  dispersion relation. We found, at least for the long and short
  wavelength modes, that $B_1>0$ and $B_2>0$ are the necessary conditions to avoid instability. 
  Furthermore, in order to avoid ghosts, one needs to set the condition $Q>0$.
\item The backgrounds for the general MGM, which are not de Sitter, if
  all the previous conditions are satisfied, at least in the scalar
  sector, will always have modes with superluminal propagations. In
  fact, for the short-wavelength ones, independently of the
  background, their propagation speed grows linearly with their wave
  number $k/a$, up to the cutoff of the theory. The presence of
  superluminal modes might be present also for the tensor modes, but
  this is only a background dependent feature, i.e.\ the propagation
  speed is independent of $k$.  Recently, in the literature,
  superluminal modes in cosmology have been discussed in the context
  of different theories, such as k-essence models. Whether or not
  these modes represent a real issue on a background which explicitly
  brakes Lorentz invariance is still matter of discussion and a clear
  and definite opinion shared by everybody on this issue is still not
  achieved
  \cite{Dubovsky:2006vk,Bonvin:2006vc,Vikman:2006hk,Hashimoto:2000ys,Babichev:2006vx,Garriga:1999vw,Bruneton:2006gf,Ellis:2007ic,Bonvin:2007mw,Babichev:2007dw,Kang:2007vs}. Some
  problems may arise from having a superluminal modes interacting with
  matter fields, either indirectly (by looking for example at the
  evolution of the clustering $\delta\rho/\rho$) or directly (such as
  the production of a kind of Cerenkov radiation), but this would
  imply to go beyond linear perturbation theory.
\item The behaviour of the scalar modes can be used in order to
  distinguish among different MGM. In fact, we have shown that
  the modified gravity models which possesses second order differential
  equations can be divided into two categories
  \begin{enumerate}
  \item General Gravity Models, for which
    $F_\sigma^2-F_\lambda\xi_\sigma\neq0$. The scalar modes
    have two degrees of freedom, but their equation of motion is non-trivial in the
    sense that it has also a $\nabla^4\Phi$ term. This terms, at least for the
    short-wavelength modes, changes the dispersion relation such that
    the group velocity of these modes becomes proportional to $k$, i.e
    $v_g=2\sqrt{B_2}k/a$ (if $B_2>0$). Therefore the theory, unless
    the vacuum is de Sitter, will always have superluminal modes. This
    is a new feature of these model which, up to our knowledge, was
    not considered before in the literature. It should be noticed that
    this feature is not a spurious one, in the sense that it can be
    gauged away. In fact, the analysis has been done through gauge
    invariant fields. Besides, no matter which gauge invariant field is
    used to study the large $k$ behaviour, exactly the same $B_2$
    coefficient appears, as the parameter in front of the $k^4$-term.
  \item Special Gravity Models, for which
    $F_\sigma^2-F_\lambda\xi_\sigma=0$. The $f(R)$ and $R+f(\GB)$
    theories belong to this class (however general $f_1(R)+f_2(\GB)$ does not),
    but they are not the only ones. In fact we have found that there
    is a larger group of theories which all share the same feature
    $B_2=0$, which distinguishes them from the General case.  This
    means that the equation of motion for the modes is similar to a
    standard wave equation, where speed of propagation is background
    dependent(except for the $f(R)$ theories for which $c^2=1$).  As
    far as we know, these are the models mostly studied in the
    literature, but actually they represent a special case of the
    whole class of MGM.
\item The de Sitter background is a ``good'' background for all
  modified gravity theories, provided that it is stable, that is if
  the scalar field has an effective positive squared mass, and
  provided that $F>0$, in order to remove ghosts degrees of freedom.
  \end{enumerate}
\end{itemize}

In this paper we have only considered the vacuum case. Although the
presence of matter is important, the vacuum itself has presented a
rich structure due to the presence of the $k^4$-term. Thinking on how
matter could change the whole picture, it is difficult to imagine that
this $k^4$-term discontinuously disappears if we add standard matter
fields (which, in the action, are expected to contribute only to the
$k^2$-terms) into the theory, because this $k^4$ dispersion relation
is solely due to the modification of gravity and not due to the nature
of matter. In this sense, the essence of the $k^4$ propagation, already
present in vacuum case, clearly supports the importance of
studying the vacuum. 

The next obvious and important extension of this work is to add matter
fields into these MGM, that we will study in another project of ours.
Although conclusive results will be given in our forthcoming paper, we
can naively expect the following effects on the propagation of modes
due to the existence of matter.  First, the master equation for the
scalar perturbation of gravity has time dependent coefficients
determined by the background dynamics.  Since the background dynamics
changes if matter fields are present, one obvious effect, by including
the matter, is that those time dependent coefficients in the master
equation will change.  Second, we are adding new degrees of
freedom. Therefore, we expect that one variable is not enough to
specify the perturbation behavior.  We will obtain at least two
coupled evolution equations for the perturbations.

\section{Conclusions}

We have studied the structure of the scalar cosmological perturbation
for the most general classes of MGM, $f(R,\GB)$ which do not have
spurious spin-2 degrees of freedom. Indeed we have discovered, for the
general case, a non-trivial propagation for the modes that has not
been studied before. In fact, in the past literature
\cite{Cartier:2001is,Hwang:2005hb}, claims were made about studying
the general class of MGM. However, we believe that this study was
never actually performed, as the authors only discussed a special
subcase of the general MGM.

This new propagation affects primarily the short-wavelength modes on
non-de Sitter backgrounds, for which their group velocity becomes
proportional to their wave number, $v_g=2\sqrt{B_2(t)}k/a$, where
$B_2$ is a background dependent quantity. This amounts to have a
non-standard dispersion relation. In turn, this implies that these
models will in general have superluminal modes. Besides we have also
given the conditions that the same modes may be stable, i.e.\ they do
not acquire a negative squared speed, or they do not become ghosts
degrees of freedom. Although there is still discussion about the
presence in cosmology of superluminal modes, we have here a new
necessary constraint to impose for these MGM, that is $0\leq
B_2<+\infty$.

However, not all MGM share this feature. Indeed there is a subclass
(to which the common $f(R)$ and $f(\GB)$ theories belong) which does
not have any longer these modified dispersion relation.

We strongly believe, that in the process to find (through a
phenomenological bottom-up approach) the ultimate low-energy effective
theory of gravity (and possibly to find at last that it coincides with
General Relativity, albeit a cosmological constant), this work can be
considered fundamental in order to set constraints to the cosmology of
these models.

The search for the gravitational action requires indeed the study of
local gravity constraints, the study of cosmological solutions, but,
also in the light of new experiments, more and more the analysis of
cosmological perturbations. It is necessary to understand the
behaviour of these same models in the presence of matter fields, but
we leave this issue for future work.

\begin{acknowledgments}
  This work is supported by the Belgian Federal Office for Scientific,
  Technical and Cultural Affairs through the Interuniversity
  Attraction Pole P6/11.
\end{acknowledgments}

\begin{appendix}
\section{Expressions of $A_i(t) $, $(i=1,...,7)$, $B_1,B_2$ and $Q$}
We give explicit forms for $A_{i}(t)$.
\begin{align}
A_1(t)&=-\frac{3(\dot F+4 H^2 \dot\xi)^2}{2 (F+4 H \dot\xi)
   [\dot F+2 H (F+6 H \dot\xi)]}\\
A_2(t)&=-\frac{12H^2\dot H}{(F+4 H \dot\xi)
   [\dot F+2 H (F+6 H \dot\xi)]}\\
A_3(t)&=-\frac{2 \dot H }{(F+4 H \dot\xi)^2}\\
A_4(t)&=-\frac{2 \{(\dot F+4 H^2 \dot\xi) [\dot F+4 (\dot H+H^2)
   \dot\xi]+F H (\dot F+8 H^2 \dot\xi)+4 H \ddot\xi
   [\dot F+H (F+8 H \dot\xi)]+F^2
   H^2\}}{(F+4 H \dot\xi) [\dot F+2 H (F+6 H
   \dot\xi)]}\\
A_5(t)&=\frac{16 H \dot\xi [\dot F+4 (\dot H+H^2) \dot\xi] (\dot F+4 H^2
   \dot\xi)+F [3 \dot F+4 (4 \dot H+3 H^2) \dot\xi] (\dot F+4
   H^2 \dot\xi)-4 \ddot\xi (\dot F+4 H^2 \dot\xi)^2}{2 (F+4 H
   \dot\xi)^2 [\dot F+2 H (F+6 H \dot\xi)]}\, \\
A_6(t)&=-\frac{2 \dot H (\dot F+4 H^2 \dot\xi)^2 \bigl\{6
   F_\sigma \bigl[\ddot H (\dot F+4 H^2 \dot\xi)+4 H \dot H [\dot F+2 (\dot H+2
   H^2) \dot\xi]\bigr]-\dot F \dot\xi\bigr\}}{3
   H [\dot F+2 H (F+6 H \dot\xi)] [2 \dot H (\dot F \dot H+2 H^2
   \dot F-48 F_\sigma H \dot H^3+8 H^4 \dot\xi)+H \ddot H (\dot F+4 H^2 \dot\xi)]}\, ,\\
A_7(t)&=\frac{2 \dot H (32 F H^3 \dot\xi+4 F^2 H^2+64
   H^4 \dot\xi^2) \bigl\{6 F_\sigma \bigl[\ddot H
   (\dot F+4 H^2 \dot\xi)+4 H \dot H
   [\dot F+2 (\dot H+2 H^2) 
   \dot\xi]\bigr]-\dot F \dot\xi\bigr\}}{3 H
   [\dot F+2 H (F+6 H \dot\xi)]
   [2 \dot H (\dot F \dot H+2 H^2 \dot F-48
   F_\sigma H \dot H^3+8 H^4 \dot\xi)+H
   \ddot H (\dot F+4 H^2 \dot\xi)]}\, .
\end{align}
A useful relation among these quantities is the following $A_2 A_6=A_1 A_3 A_7$, moreover both $A_6$ and $A_7$ are proportional to $F_\sigma^2-F_\lambda \xi_\sigma$, so that they identically vanish for the special cases.
We also have
\begin{equation}
Q=\frac{A_1\,(F+4H\dot\xi)^2}{2(A_2A_7-1)[\dot F+2 H (F+6 H \dot\xi)]}\, .
\end{equation}
Also we have defined
\begin{align}
  B_1&=\frac{-A_6 \dot A_2-A_2^2 (A_4
   A_6 A_7-A_7 \dot A_6+A_6
   \dot A_7)+A_2 (A_4
   A_6-\dot A_6)+A_5 (A_2
   A_7-1)^2}{A_1 (A_2 A_7-1)}\, ,\\
  B_2&=-\frac{A_2\,A_6^2}{A_1^2\,A_7}=
\frac{64 \dot H^2 \left(F_\sigma^2-F_\lambda \xi_\sigma\right)}{3
   [F_\lambda+8 H^2 (F_\sigma+2 H^2 \xi_\sigma)]
   \bigl\{F+24 H \bigl[\ddot H (F_\sigma+4 H^2 \xi_\sigma)+4
   H \dot H [F_\sigma+2 \xi_\sigma (\dot H+2
   H^2)]\bigr]\bigr\}}\, .
\end{align}

\end{appendix}

\bibliography{k4bib.bib}

\end{document}